\documentclass[prb,showpacs,twocolumn,preprintnumbers,amsmath,amssymb,floatfix]{revtex4}
\usepackage[dvips]{graphicx}
\usepackage[latin1]{inputenc}
\begin{document}
\draft

\newcommand{\cu} {$^{63}$Cu }
\newcommand{\cuiso} {$^{63,65}$Cu }
\newcommand{\etal} {{\it et al.} }
\newcommand{\ie} {{\it i.e.} }
\newcommand{\aucr}{CeCu$_{5.9}$Au$_{0.1}$ }
\newcommand{\auaf}{CeCu$_{5.2}$Au$_{0.8}$ }
\newcommand{\aux}{CeCu$_{6-x}$Au$_{x}$ }
\newcommand{\ip}{${\cal A}^2$ }

\hyphenation{a-long}

\title{Anomaly in YBa$_2$Cu$_4$O$_8$ charge distribution below T$_c$: a zero-field $\mu$SR study}

\author{P. Carretta$^1$, A. Keren $^{2,3}$, J.S. Lord$^3$, I. Zucca$^1$, S.M.Kazakov$^4$ and J. Karpinski$^4$}
\affiliation{$^1$Department of Physics ``A.Volta" and Unit\`a
INFM, University of Pavia, Via Bassi 6, I-27100, Pavia (Italy)}
\affiliation{$^2$ Department of Physics, Technion - Israel
Institute of Technology, Haifa 32000 (Israel)}
 \affiliation{$^3$
ISIS Facility, Rutherford Appleton Laboratory, Chilton OX11 0QX,
(United Kingdom) } \affiliation{$^4$Solid State Physics Laboratory
ETH, 8093 Zürich, Switzerland}
\widetext

\begin{abstract}

Zero-field $\mu$SR measurements in $^{63}$Cu isotope enriched and
natural YBa$_2$Cu$_4$O$_8$ powders are presented. The temperature
dependence of the  $\mu^+$ relaxation rate is characterized by a
sizeable enhancement below T$_c$. The comparison of the asymmetry
decay in the two samples reveals that the $\mu^+$ relaxation is
driven by nuclear dipole interaction from 300 K down to 4.2 K. It
is argued that the increase in the relaxation below T$_c$
originates from a change in $\mu^+$ site, possibly due to a
modification in the charge distribution within CuO chains.

\end{abstract}

\pacs {76.75.+i, 74.72.-h, 71.45.Lr} \maketitle

\narrowtext

\section{Introduction}

In strongly correlated metals the electrons can be subject to
interactions which are of the order of a fraction of the bandwidth
and can cause an enhancement of instabilities, as
superconductivity, and to crossovers among different regimes
\cite{Fabr}. One of the most intriguing examples of such a
scenario is the phase diagram of the high T$_{c}$ superconductors,
which are characterized by a relatively narrow bandwidth, when
compared to conventional superconductors, and by sizeable exchange
and local lattice interactions. Although it has been widely
explored experimentally and several crossover temperatures
evidenced, the microscopic mechanisms leading to such a complex
phase diagram are still unexplained. The origin of the pseudo-gap
is not yet clear. The occurrence of a mesoscopic phase-separation
and/or of a charge order in the CuO$_{2}$ planes over a wide
doping range are still subject of an intense scientific debate.
This debate centers around the relevence of such phase-separation
to the mechanism of superconductivity. Moreover, it is not
established which is the role of CuO chains on the electronic
properties of certain families of cuprates. Namely, if CuO chains
can be considered to a certain extent decoupled from the
underlaying CuO$_{2}$ planes and be characterized by an
independent phenomenology, typical of a Tomonaga-Luttinger liquid
\cite{Tomo}. In these circumstances, the observation and
clarification of every phase transition/crossover could be very
useful.

Few years ago Kramers and Mehring have revealed a new crossover
temperature below T$_c$ of optimally doped YBa$_2$Cu$_3$O$_{6+x}$
(Y123). \cite{kramer} They observed a peak in the Cu(2) nuclear
transverse relaxation rate, with a concomitant increase in the NQR
linewidth. Hence, they suggested that these anomalies could
possibly be associated with the onset of a charge order
below $T\simeq 40$ K in the CuO$_2$ plane. Later Sonier et al. %
\onlinecite{Sonier1}, have observed an anomalous increase in the
zero-field (ZF) $\mu^+$ relaxation rate below T$_c$ in the same
compound and, at first,
associated it with the pseudo-gap crossover temperature T$^*$. Later on \cite%
{Sonier2}, it was realized that this anomalous increase occurred
at the same crossover temperature detected by NQR suggesting that
both techniques are detecting the same crossover temperature.

In order to clarify these aspects, namely if the increase in the
ZF $\mu$SR relaxation occurs at T$^*$, if it originates from a
modification in the local field distribution due to electron or
nuclear spins and if it is
related to a charge order, we have performed ZF $\mu$SR measurements in $%
^{63}$Cu enriched and natural YBa$_2$Cu$_4$O$_8$ (Y124). The
peculiarity of Y124 is that, unlike Y123, it is characterized by a
well defined oxygen stoichiometry and by a precise value of
T$^*\simeq 150$ K \cite{Mali} and hence it is the best system
where one can check if there is any correlation between T$^*$ and
the anomalous increase in the muon relaxation rate. Moreover, the
comparison of $\mu^+$ relaxation rate in samples with different
abundances of $^{63}$Cu isotope allows to clarify the origin of
the local field distribution at muon sites.

\begin{figure}[t!]
\vspace{30mm} \hspace{75mm}
\includegraphics[scale=0.4]{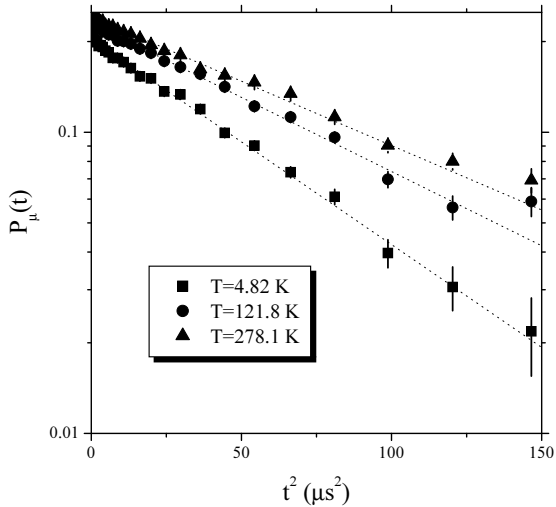}
\vspace{-40mm} \caption{Decay of the ZF $\mu^+$ polarization in
Y124 powder samples with natural abundant $^{63}$Cu at a few
selected temperatures. The decay is reported in semi-log scale
versus $t^2$ in order to evidence the accuracy of the fit (dotted
lines) according to Eq.1.} \label{XRD}
\end{figure}

\section{Experimental results}

Y124 powder samples were grown following a standard procedure
\cite{Karp}. Appropriate amounts of Y$_2$O$_3$ (99.99\% Alfa
Aesar), BaCO$_3$ (99.98\%, Aldrich), enriched Cu$^{63}$O (99.9\%)
and natural CuO (99.99\%, Aldrich) were mixed then pressed into
pellets and annealed in air at 850 - 910ºC for 150 h, with several
intermediate regrinding. X-ray diffraction revealed that resulting
samples were mixture of R-123 and CuO. These samples were placed
into Al$_2$O$_3$ - crucibles and subjected to the high oxygen
treatment in a double-chamber high-pressure system. The
temperature was first raised to 1000C at 10/min, and was held at
this temperature for 60 h, followed by cooling to room temperature
at 5 /min. The value of oxygen pressure was kept at 480 bar.

ZF $\mu$SR measurements were performed at ISIS pulsed source on
MUSR beam line, using $29$ MeV/c spin-polarized muons. The use of
an intense pulsed muon source as ISIS has the major advantage that
it allows to measure slow relaxation rates with the highest
accuracy. The background signal due to the cryostat and sample
holder was estimated from the slowly decaying part of the
polarization in low-temperature transverse field measurements.
During the ZF measurements an automatic compensation of the
magnetic field was active in order to grant a magnetic field on
the sample below a few tenths of mGauss. This is important in
order to assure that no extra-contribution to the relaxation
associated with the trapping of the magnetic flux is present.

In Fig. 1 the ZF decay of the muon polarization for the $^{63}$Cu
enriched sample is reported vs. $t^2$ for a few selected
temperatures. As one can notice, the form of the decay law is
Gaussian below 12 $\mu$s and  does not change upon cooling from
about 200 K down to 4.2 K. In fact, the data can be nicely fit
with a static Gaussian Kubo-Toyabe function \cite{KT,Uem}, the one
theoretically expected when the relaxation is driven by nuclear
moments,
\begin{equation}
\label{eq:1}
          P_{\mu}(t)={1\over 3} + {2\over 3}(1-
          \sigma^2t^2)exp(-{1\over 2}\sigma^2t^2) ,
\end{equation}
where $\sigma = \gamma\sqrt{<\Delta h^2>}$, with $\gamma=
2\pi\times 13.55$ kHz/Gauss the muon gyromagnetic ratio and
$\sqrt{<\Delta h^2>}$ the amplitude of the local field
distribution experienced by the muons. To second order in time $t$
this function is identical to a Gaussian and the asymmetry decay
plotted as $log  P_{\mu}(t)$ vs. $t^2$ (Fig.1) is given by a
straight line. The decay of the muon polarization was observed to
be faster in the sample with natural isotope abundance with
respect to the $^{63}$Cu enriched samples (Fig. 2). The values
derived for $\sigma$ from Eq.\ref{eq:1} for both samples in the
300 K to 4.2 K temperature range are finally reported in Fig. 3. A
small decrease in the relaxation with increasing temperature is
observed around $200$ K and possibly associated with $\mu^+$
diffusion \cite{diffuse}. One notices that no anomaly is observed
at T$^*\simeq 150$ K, definitely ruling out the occurrence of any
increase in $\sigma$ at T$^*$. However, a pronounced increase is
clearly visible below T$_c$ and is quantitatively similar to the
one observed in optimally doped Y123 by Sonier et al.
\cite{Sonier1} for $T\simeq 40$ K.

\begin{figure}[t!]
\vspace{30mm} \hspace{75mm}
\includegraphics[scale=0.4]{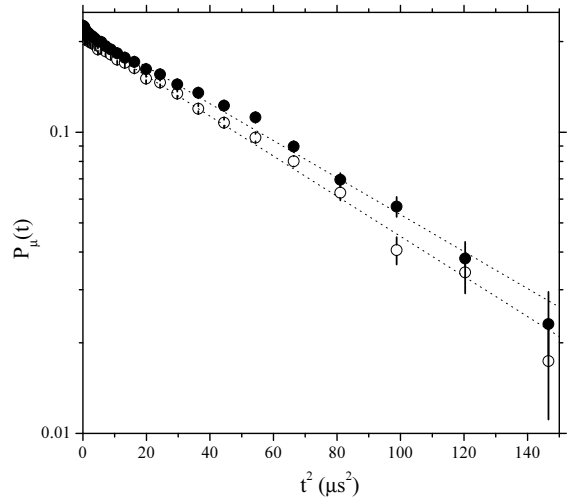}
\vspace{-40mm} \caption{Decay of the ZF $\mu^+$ polarization at
T$\simeq 20$ K for the $^{63}$Cu enriched (closed circles) and not
enriched (open circles) Y124 powder samples. The decay is reported
in semi-log scale versus $t^2$ in order to evidence the accuracy
of the fit with Eq.1.} \label{XRD}
\end{figure}

\section{Discussion and Conclusion}

The dependence of $\sigma$ on the $^{63}$Cu isotope abundance is a
straightforward evidence that the ZF muon relaxation in Y124 is
driven by the interaction with nuclear magnetic moments, over all
the explored temperature range. In fact, taking into account that
the natural abundance is given by 69\% of $^{63}$Cu
($\gamma_{63}/2\pi= 11.285$ MHz/Tesla) and 31\% of $^{65}$Cu
($\gamma_{65}/2\pi= 12.089$ MHz/Tesla), the ratio of the second
moment of the field distributions due to nuclear dipolar
interaction in the enriched and natural samples scales as
\cite{Schenck}
\begin{equation}
\label{eq:2}
          {\sigma_{nat}^2\over \sigma_{e}^2}=
          (0.31*{\gamma_{65}^2\over \gamma_{63}^2}+0.69)= 1.046 ,
\end{equation}
with $\sigma_{nat}$ and $\sigma_{e}$ the relaxation rates for the
natural and enriched samples, respectively. One observes in Fig. 3
that the relaxation rate of the natural sample is slightly larger
than the one of the isotope enriched sample, as expected from
Eq.\ref{eq:2}. In Fig.4 the ratio $\sigma_{nat}/\sigma_{e}$ is
reported for a few selected temperatures at which measurements in
both samples were performed and a reasonable agreement with the
ratio expected on the basis of nuclear dipole interaction is
found.

\begin{figure}[t!]
\vspace{40mm} \hspace{75mm}
\includegraphics[scale=0.49]{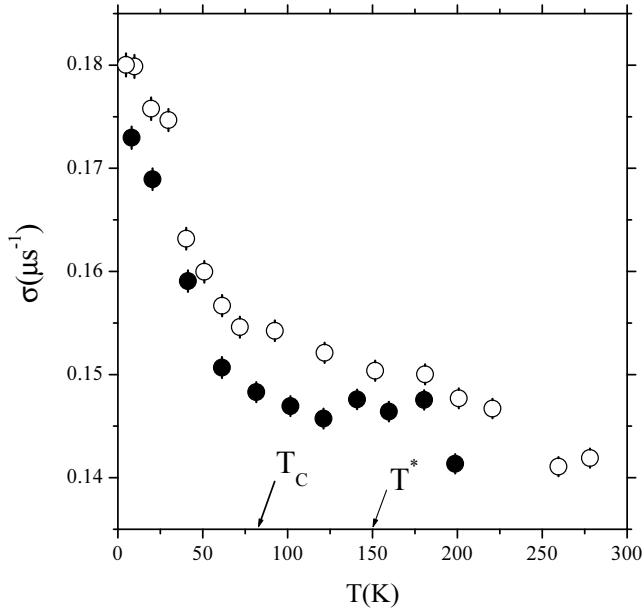}
\vspace{-40mm} \caption{Temperature dependence of the ZF $\mu$SR
relaxation rate $\sigma$ (see Eq. 1) in isotope enriched (closed
circles) and not-enriched (open circles) Y124 powder samples. The
dotted line shows the expected behavior of $\sigma$ in the
non-enriched sample once the local field distribution is rescaled
by the gyromagnetic ratio and the isotope abundances (see text). }
\label{XRD}
\end{figure}

In principle one could associate the increase in $\sigma$ below 40
K with a crossover from a high temperature regime, where $\mu^+$
diffusion occurs, to a low temperature one where the muon is
localized, as observed in several metals \cite{Schenck}. However,
this hypothesis is in conflict with the observation that the decay
of the muon polarization is nicely fit with a static Kubo-Toyabe
function over all the explored temperature range. Hence the
increase in the decay of the muon polarization is not associated
with a slowing down of the muon dynamics but rather to a
modification in the field distribution probed by the muons.

\begin{figure}[t!]
\vspace{30mm} \hspace{75mm}
\includegraphics[scale=0.4]{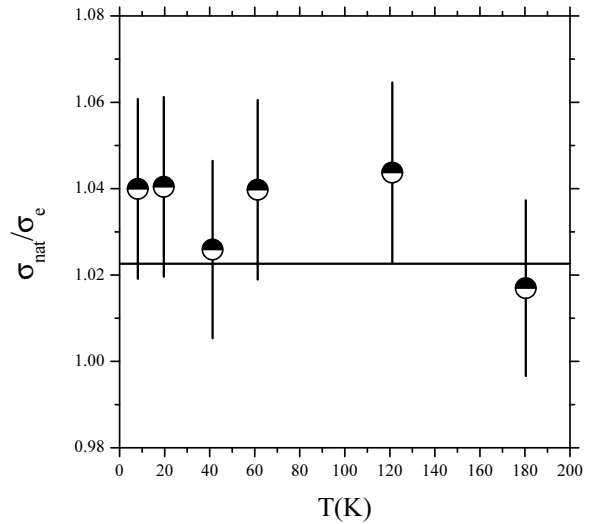}
\vspace{-40mm} \caption{Ratio of the ZF $\mu$SR relaxation rate
$\sigma$  in isotope enriched  and not-enriched Y124 powder
samples. The solid line shows the theoretical value for this ratio
calculated according to Eq.2. } \label{XRD}
\end{figure}

Such a change can occur only if the relative position between the
muon and the nuclei changes, namely if $\mu^+$ site changes. A
modification in the $\mu^+$ site occurs as a consequence of a
variation in the crystal field, associated with a modification in
the surrounding charge distribution. Also Sonier et
al.\onlinecite{Sonier2} have suggested a similar scenario for
optimally doped Y123 after reconsidering the interpretation of
their data.

In principle, one could associate the increase in $\sigma$ with a
lattice distortion. However, in the cuprates, although clear signs
of a modification in the phonon spectra have been detected
\cite{Egami} suggesting a strong electron-phonon coupling, no
structural distortions have been observed at T$_c$. Moreover, it
has to be remarked that in optimally doped Y123 the analogous
increase in $\sigma$ is detected well below T$_c$, pointing out
that it cannot be directly related to the superconducting
transition. On the other hand, a change in the $\mu^+$ site could
arise from a modification in the charge distribution. In
PrBa$_2$Cu$_3$O$_7$ and Y123 Grevin et al.
\onlinecite{grevin1,grevin2} have suggested on the basis of NQR
measurements that a CDW order might set in. In Y124 the NQR
results  do not seem to support unambiguously such a scenario,
although a clear anomaly in the NQR frequency of the chain copper
was revealed at T$_c$ \cite{Raffa} suggesting a modification in
the charge distribution within the chains. It is worth to mention
that such anomalies do involve the chains as they are observed
only in compounds with completely filled CuO chains, as optimally
doped Y123 and Y124, and they are absent in compounds without
chains as La$_{2-x}$Sr$_x$CuO$_4$ \cite{LacuO}. Therefore, the
increase in the $\mu$SR relaxation rate at low-temperature in Y124
and optimally doped Y123 seems to signal a crossover from a high
temperature disordered arrangement of the charge distribution to a
regime where at least short range correlations in the charge
density set in within the CuO chains \cite{Sendyka}. Hence, it
appears that the CuO chains in Y123 and Y124 could be
characterized by an independent phenomenology, as if they were
almost decoupled from the superconducting CuO$_2$ layers. This
hypothesis is also supported by the observation of coexisting
magnetic order and superconductivity in the adjacent CuO$_2$ and
RuO layers of ruthenocuprate superconductors \cite{ruteno}.

In conclusion, from a careful analysis of the ZF $\mu$SR
relaxation in isotope enriched and natural Y124 powders we have
observed an anomalous increase in the relaxation rate below T$_c$
which has to be unambiguously associated with a change of $\mu^+$
site. This modification suggests a common scenario for optimally
doped Y123 and Y124, with a crossover to a low-temperature regime
where the charge distribution within the CuO chains varies,
possibly due to the growing charge density correlations.

\section*{Acknowledgement}

The research activity in was supported by the Italian project
MIUR-FIRB {\it Microsistemi basati su materiali magnetici
innovativi strutturati su scalla nanoscopica}. A. Keren's activity
was supported by the Israeli Science Foundation.




\end{document}